\documentclass[jmp,twocolumn,showpacs,superscriptaddress,nofootinbib]{revtex4-1}

\usepackage{amsfonts}
\usepackage{amsmath}
\usepackage{amssymb}
\usepackage{upgreek}
\usepackage{bm}
\usepackage{dcolumn}
\usepackage{epsfig}
\usepackage{graphicx}
\usepackage{graphics}
\usepackage[latin1]{inputenc}
\usepackage{latexsym}
\usepackage{rotating}
\usepackage{hyperref}
\usepackage{xspace} 
\usepackage[usenames]{color}

\usepackage{ulem}
\definecolor {darkgreen}{rgb}{0.2,0.7,0.2}

\newcommand\be{\begin{equation}}
\newcommand\ba{\begin{eqnarray}}
\newcommand\ee{\end{equation}}
\newcommand\ea{\end{eqnarray}}

\begin{document}
\title{Resonances in Extreme Mass-Ratio Inspirals: \\
Asymptotic and Hyperasymptotic Analysis}

\author{Jonathan Gair}
\affiliation{Institute of Astronomy, Madingley Road, Cambridge, CB30HA, United
Kingdom}

\author{Nicol\'as Yunes}
\affiliation{Department of Physics, Montana State University, Bozeman,
Montana 59717, USA}
\affiliation{MIT and Kavli Institute, Cambridge, Massachusetts 02139, USA}

\author{Carl M.~Bender}
\affiliation{Department of Physics, Washington University, St.~Louis, Missouri
63130, USA}

\date{\today}

\begin{abstract} 
An expected source of gravitational waves for future detectors in space are the
inspirals of small compact objects into much more massive black holes. These
sources have the potential to provide a wealth of information about astronomy
and fundamental physics. On short timescales the orbit of the small object is
approximately geodesic. Generic geodesics for a Kerr black hole spacetime have a
complete set of integrals and can be characterized by three frequencies of the
motion. Over the course of an inspiral, a typical system will pass through
resonances where two of these frequencies become commensurate. The effect of the
resonance will be to alter significantly the rate of inspiral for the duration
of the resonance. Understanding the impact of these resonances on gravitational
wave phasing is important to detect and exploit these signals for astrophysics
and fundamental physics. Two differential equations that might describe the
passage of an inspiral through such a resonance are investigated. These differ
depending on whether it is the phase or the frequency components of a Fourier
expansion of the motion that are taken to be continuous through the resonance.
Asymptotic and hyperasymptotic analysis are used to find the late-time analytic
behavior of the solution for a system that has passed through a resonance.
Linearly growing (weak resonances) or linearly decaying (strong resonances)
solutions are found depending on the strength of the resonance. In the
weak-resonance case, frequency resonances leave an imprint (a resonant memory)
on the gravitational frequency evolution. The transition between weak and strong
resonances is characterized by a square-root singularity, and as one approaches
this transition from above, the solutions to the frequency resonance equation
bunch up into families exponentially fast. 
\end{abstract}

\maketitle

\section{Introduction}
\label{intro}

``{\emph{Divergent series are the invention of the devil, and it is shameful to
base on them any demonstration whatsoever.}}'' Niels Hendrik Abel's 1828
statement~\cite{Boyd99thedevil's} suggests that asymptotic analysis, which
commonly leads to divergent series, should not be applied to problems of
physical interest. Asymptotics, however, has become an invaluable tool for
physicists seeking approximate analytic solutions. Multiple-scale analysis,
which includes boundary-layer theory and WKB theory, allows us to understand
diverse problems, such as semiclassical quantum theory, airplane wing design and
turbines~\cite{Bender}. In the context of general relativity, asymptotic
(post-Newtonian) series~\cite{Yunes:2008tw,Zhang:2011vh} constitute the basis of
the filters used in current gravitational wave detectors to extract signals from
the noise.

Resonances are a common occurrence in physical phenomena. In a traditional
oscillatory system, a resonance is a point in frequency space where the system
stores and transfers energy between kinetic and potential modes, allowing a
small driving force to generate large amplitude oscillations. Resonant phenomena
can occur in many vibrational or wave-like systems and include electromagnetic
resonances, nuclear magnetic resonance, electron spin resonance, and so on. In
the realm of general relativity, black holes can sometimes be treated as
resonators, as they relax after being perturbed (see, for example,
\cite{Berti:2009kk}). 

In the context of general-relativistic orbital mechanics, {\it resonance} has
recently been adopted to represent a slightly different phenomenon, namely, the
enhancement of gravitational-wave energy dissipation due to the lack of
cancellation of oscillatory modes that for generic inspirals average out
\cite{Flanagan:2010cd,Gair:2010iv}. This is particularly relevant for extreme
mass-ratio inspirals (EMRIs), in which a small compact object, such as a
stellar-mass black hole or a neutron star, orbits around a supermassive black
hole \cite{2007CQGra..24..113A}. In such a two-body system the smaller object
slowly spirals inwards due to gravitational-wave energy-momentum losses, on a
radiation-reaction timescale much longer than the orbital one. This inspiral is
usually modeled by computing an orbit-average of the gravitational-wave energy
flux. This procedure discards terms proportional to odd-powers of sines or
cosines of the sum of the orbital phases
\cite{Poisson:1993vp,Hughes:1999bq,Hughes:2001jr,Pound:2007th}. For orbital
configurations or points in frequency space at which the sum of the orbital
frequencies vanishes, the orbit-averaged energy flux is not equal to the limit
of the orbit-averaged fluxes for nearby, nonresonant orbits. This is because
harmonics of the frequency that vanish on resonance contribute to the secular
component of the change in the orbital elements on resonance
~\cite{Flanagan:2010cd,Gair:2010iv}, but average to zero for off-resonance
orbits. Unlike traditional oscillators, however, there is no {\emph{external}}
driving force in the EMRI case; the emitted gravitational waves drive the
inspiral themselves and the resonance is caused by the orbital frequencies
becoming commensurate, that is, some linear combination of the three frequencies
with integer coefficients vanishes at resonance.

A secularly growing radiation-reaction force can leave strong imprints on the
orbital motion, even if this secular growth is active for a very short time. 
These imprints can then propagate into the gravitational waves emitted and could
have important consequences for gravitational-wave detection. Unlike conventional telescopes operating in the electromagnetic spectrum, current gravitational-wave detectors will not observe
signals above the average noise. Instead, signals are expected to be buried deep
in the noise, and will be extracted using filters based on the 
expected signals. Although EMRIs are not expected to be detected with current
ground-based gravitational wave detectors, they are a key target for future
space-based detectors for which accurate EMRI filters will be needed. It is
therefore important to understand how EMRI resonances can affect the emitted
gravitational waves. 

\subsection*{Resonances in Extreme Mass-Ratio Inspirals}
The extreme-mass-ratio (typically $10^{-6}$ -- $10^{-5}$) ensures that over
short timescales the orbit of the smaller object in an EMRI system is
approximately geodesic. It is therefore appropriate to use an
``osculating-element'' formulation in which the EMRI is identified by a sequence
of geodesics~\cite{Pound:2007th,Gair:2010iv}. Geodesics in a Kerr background are
uniquely characterized by three constants of the motion, energy $E$,
$z$-component of angular momentum $L_z$, and {\it Carter constant} $Q$, and four
initial phases that specify the coordinates of the object at a particular time.
The position and velocity of an object uniquely identifies a geodesic. Since the
evolution of the orbit is governed by a second-order differential equation, the
values of these seven geodesic parameters at each point on the inspiral provide
an alternative parametrization of the inspiral. The time-evolution of the
geodesic constants of the motion is 
\be
\frac{dJ_{\nu}}{dt}=\epsilon{\cal{F}}^{\rm SF}_\nu(q,J)+{\cal{O}}(\epsilon^2)\,,
\label{J-eq}
\ee
where $J_{\nu}=(E,L_z,Q)$ is a vector of these constants,
while $\epsilon$ is the mass ratio, $q$ is an angle phase variable, and ${\cal{F
}}^{\rm SF}_\nu$ is the ``self-force''. The rate of change of $J_\nu$ can then be
used to construct the rate of change of the orbital frequencies in a similar
form.

For any given geodesic, the self-force can be expanded in a Fourier series in terms of the fundamental
orbital frequencies. These frequencies can be mapped to
the geodesic constants of the motion. A geodesic resonance occurs when the ratio
of the frequencies of the radial and vertical motion is a rational number. The third frequency, that of meridional motion, is not relevant for resonances due to the axisymmetry for the background Kerr spacetime. Henceforth, we only consider the dependence of the self-force on
the two frequencies that can lead to a resonance. 

Let us then expand the self-force in a two-frequency Fourier series, where one
of the frequencies ($\omega$) approaches zero while the other ($f$) remains
finite:
\be
\frac{d\omega}{dt}=\epsilon\sum_{\ell,n}G_{\ell n}\cos\left[\left(\ell\omega+nf
\right)t\right]+H_{\ell n}\sin\left[\left(\ell\omega+nf\right)t\right]\,,
\label{w-eq}
\ee
where $G_{\ell n}$ and $H_{\ell n}$ are time-independent Fourier coefficients
that depend on the orbital parameters. Clearly, when $\ell\omega+nf\neq0$, the
cosine and sine terms average out for sufficiently long integration times.
However, at resonance, where $\ell\omega=-nf$ (which in this case we take to be
$n=\omega=0$), the cosine function goes to unity, leaving a sum of secular
(zero-frequency) Fourier coefficients. 

Let us now further assume that the Fourier coefficients $(G_{\ell n}, H_{\ell n}
)$ vary smoothly as the resonance is approached, such that they can be expanded
as their on-resonance values plus a correction of ${\cal{O}}(\epsilon^{1/2})$.
Such corrections can always be made small by choosing a sufficiently small
$\epsilon$, independent of the magnitude of $\omega$. Similarly, corrections
from other terms of ${\cal{O}}(\epsilon)$ on the right side of \eqref{J-eq} can
be ignored. 

We are now left with a number of rapidly oscillating terms (those with $\ell
\omega+nf\neq 0$) and also terms that slowly oscillate away from resonance and
then vanish at resonance. The rapidly oscillating terms are less important
because they average to zero on a short timescale. Changing variables to $y
\equiv\omega/\sqrt{\epsilon}$ and $x\equiv\sqrt{\epsilon} t$, we find that
Eq.~\eqref{w-eq} becomes
\be
\frac{dy}{dx}=\sum_{\ell n}G_{\ell n}\cos\left(\ell xy+\frac{nfx}{\sqrt{\epsilon
}}\right)+H_{\ell n}\sin\left(\ell xy+\frac{n f x}{\sqrt{\epsilon}}\right)\,.
\label{probdef}
\ee
Expanding the sum for the first few $(\ell,n)$ modes and rescaling by $\tilde{y}
=y/\sqrt{G_{00}}$, $\tilde{x}=\sqrt{G_{00}} x$, we then have 
\be
y'=1+k\cos\left(x\,y\right)\,,
\label{diffeq}
\ee
where we have dropped the tildes, prime denotes differentiation with respect to
$x$, and, in principle, the parameter $k\equiv G_{01}/G_{00}$ is known. In most
scenarios $k\ll 1$, but there could be orbits for which $k={\cal{O}}(1)$. 

In deriving \eqref{diffeq}, we have made several approximations: (a) we have
ignored the rapidly oscillating $\epsilon^{-1/2}$ terms, and thus considered
only the $n=0$ modes; (b) we have ignored a phase constant induced by the $H_{10
}$ term; (c) we have considered only the $\ell=1$ mode because these are the
dominant ones and are on resonance for the longest time. Assumption (a) is
justified, given that the rapidly oscillating components tend to average out
over a sufficiently long integration time. The relaxation of assumptions (b) and
(c) will be addressed more carefully in Sec.~\ref{gen-res}. 

In addition to the approximations described above, \eqref{diffeq} also makes the
critical assumption that the self-force can be expanded as a Fourier series in
the frequency with argument $(\ell\omega+nf) t$ and that the coefficients of
this expansion are continuous at resonance. An alternative way to write the same
equation off resonance would be as an expansion in the phase with argument $\ell
\phi+n\psi$, where $(\phi,\psi)$ are angle variables. For a geodesic, the time derivatives of
$(\phi,\psi)$ are the frequencies $(\phi',\psi')\equiv(\omega,f)$, but if one
regards these phase angles as fundamental and assumes that the coefficients of
that expansion are continuous at resonance, one ends up with a slightly
different differential equation:
\be
\phi''=1+k\cos{\phi}\,,
\label{diffeq2}
\ee
which admits the first integral
\be
\frac{1}{2}\left(\phi'\right)^2=\phi+k\sin{\phi}+\phi'(0)\,,
\ee
where $\phi'(0)$ is an integration constant. 

Equations~\eqref{diffeq} and~\eqref{diffeq2} give two alternative descriptions
of an EMRI resonance, but they are not equivalent. To make this clear, we
rewrite \eqref{diffeq2} in terms of $y$:
\be
y'=1+k\cos\left(\int y dx\right)\,.
\label{diffeq3}
\ee
This equation is equivalent to \eqref{diffeq} only in the limit $x y'\ll 1$. In
this paper we seek solutions to these two differential equations in the limit $x\to
+\infty$. The equations are deceptively simple (they are just ordinary differential equations) but due to the nonlinearity, finding exact solutions is impossible.

This paper describes the solution to both the ``frequency-resonance''
differential equation \eqref{diffeq} and the ``phase-resonance'' differential
equation \eqref{diffeq2} and is organised as follows: The leading-order behavior
of the solution at late times ($x\to\infty$) for both frequency and phase
resonances are calculated in Sec.~\ref{sec:LO}. The higher-order behavior in $k$
of these solutions is given in Sec.~\ref{higher-order}. Section~\ref{transition}
describes the qualitative change in behavior as $k$ transitions from $k>1$ to
$k<1$. Section~\ref{gen-res} discusses generalisations of the resonance equations and explains
how the solutions are modified. Section~\ref{conclusions} gives some conclusions
and describes possible future work. 

\section{Leading-Order Asymptotic Behavior for Large $x$}
\label{sec:LO}

Equations~\eqref{diffeq} and~\eqref{diffeq3} describe simple models of nonlinear
resonant behavior and similar versions of these equations have been studied
before. In fact, equations of the form $y'=f(\alpha x+\beta y)$ or $y'=f(y/x)$
have solutions in quadrature. However, equations of the form $y'=f(xy)$ cannot
be solved exactly. Instead, one relies on asymptotic techniques to understand
their behavior.

The prototypical equation to study with these tools is \cite{Bender} 
\be
y'=\cos{xy}\,,
\label{simpler-ODE}
\ee
whose asymptotic expansion in the limit $x\to+\infty$ is
\be
y(x)\sim\frac{a}{x}\qquad(x\to+\infty),
\ee
where $a=(n+1/2)\pi$ and $n$ is an integer. For a slowly varying solution, as
$x\to+\infty$, $y'\ll 1$, which implies that $\cos{xy}\ll 1$ and thus $xy=a\sim
(n+1/2)\pi$. In fact, one can show that corrections to this asymptotic solution
scale with powers of $(1/x)^{m}$ for $m>1$~\cite{Bender}. Similar techniques can
be used to show that the solution to equation $y'=\tan{2xy}$ also behaves as
$y\sim(2n+1)\pi/x$ as $x\to+\infty$ \cite{Bender:2009jg,Bender:2009wx}.

\subsection{Frequency Resonances}
\label{lead-order-sec}

Let us first consider the case $k>1$ of \eqref{diffeq}. For slowly varying
solutions $y'\ll 1$, $1+k\cos{xy}\ll 1$ and then 
\be
y\sim\frac{\arccos(-1/k)}{x}\,\qquad{\rm{as}}\quad x\to\infty,k>1.
\ee 
When $k<1$, the above solution does not exist and more subtle asymptotic
techniques must be used.

Next, we consider the $k<1$ case. The form of \eqref{diffeq} suggests that $y
\sim y_{c}\equiv ax$ as $x\to+\infty$. Let us then try this ansatz, which when
inserted in \eqref{diffeq} gives
\be
a=1+\cos(ax^2)\,.
\ee
Clearly, the $y_c$ ansatz is not a proper solution. We can understand this by
averaging the cosine term over all $x$ 
\be
a=1+\left<\cos(a x^2)\right>\,,
\ee
where the angle brackets stand for averaging. The Fresnel cosine
function is defined via the integral 
\be
C(x)\equiv\left(\frac{2 a}{\pi}\right)^{1/2}\int_0^x \cos(a x'^2)dx'\,. 
\label{fresCdef}
\ee
and as $x\to+\infty$, $C(x)\sim1/2+\sin(\pi x^2/2)/(\pi x)$. At very large $x$ then, 
\be
\int\cos(a x^2)dx\sim\frac{1}{4}\left(\frac{2\pi}{a}\right)^{1/2}+\frac{\sin(a
x^2)}{2 a x}\,.
\ee
Note, however, that the second term still depends on $x$, so
our ansatz $y_{c}$ is still not a valid solution. 

These considerations motivate the improved ansatz $y\sim y_1$ as $x\to+\infty$,
where
\be
y_1\equiv(1+c)x+b+\frac{a_1}{x} \sin[(1+c)x^2 + bx]\,
\label{freqLOansatz}
\ee
with constants $a_{1}$, $b$, $c$ to be determined. Inserting this ansatz into
\eqref{diffeq}, we obtain
\ba
\label{preliminary-step}
&&c+2(1+c)a_1\cos[(1+c)x^2+bx]+{\cal{O}}\left(x^{-1}\right)\\ \nonumber &\sim&
k\cos[(1+c)x^2+bx]\,\cos\{a_1 \sin[(1+c)x^2 + bx)]\}\nonumber\\ \nonumber 
&-& k \sin[(1+c)x^2 + bx] \sin\{a_1 \sin[(1+c)x^2 + bx]\}\,,
\ea
where we have expanded the cosine function with standard trigonometric
identities. Comparing terms of the left and right sides of this equation, we see
that $a_1={\cal{O}}(k)$. 

Let us now require that $k\ll1$. Since the sine and cosine functions are bounded
by unity and since $a_1={\cal{O}}(k)\ll1$, we know that $a_1\sin[(1+c)x^2+bx]
\ll1$ for all $x$, and we can therefore expand the cosine and sine
functions on the righthand side of \eqref{preliminary-step}. Performing the
expansion, we find that \eqref{preliminary-step} becomes
\begin{align}
\label{preliminary-step2}
c+2(1+c)a_1\cos[(1+c&)x^2 + bx]\sim k\cos[(1+c)x^2+bx]\nonumber \\ 
&- a_1 k \sin^{2}[(1+c)x^2 + bx]
\end{align}
to ${\cal{O}}(x^{-1},k^3)$ from which we infer that $a_1 \sim k/[2(1+c)]$. We
are then left with 
\be
\label{preliminary-step3}
c\sim - \frac{a_1 k}{2} \left\{1 - \cos[2 (1+c)x^2 + bx] \right\}\,,
\ee
which implies that $c\sim-ka_{1}/2$, and thus, $c\sim-k^{2}/4$. Substituting
this back into $a_1$, we find that $a_1\sim k/2$. The second term of
\eqref{preliminary-step3} is not included in $c$ because it must be canceled by
terms of ${\cal{O}}(k^2)$ in $y$, which we have neglected here. Our solution to
(4) then becomes
\be
y_1=b+\left(1-\frac{k^2}{4}\right)x+\frac{k}{2x}\sin\left[\left(1-\frac{k^{2}}
{4} \right)x^2 + bx \right]\,, \label{first-order-y}
\ee
with remainders of ${\cal{O}}(x^{-1},k^{3})$, and where $b$ remains undetermined
and depends on the initial conditions. We have solved \eqref{diffeq} numerically
in the range $k\in(0,0.5)$ and $x\in(0,10^{3})$ and verified that indeed
\eqref{first-order-y} is a good approximation to the numerical solution, as we
show in Sec.~\ref{higher-order}.

The frequency evolution described by \eqref{first-order-y} is particularly
interesting. At late times, the behavior of the frequency is dominated by the
term linearly proportional to $x$, with all others becoming subdominant. The
slope of the frequency, however, is dependent on $k$. That is, as the
physical system goes through a resonance, it acquires a slope correction that
depends on the properties of the resonance (that is, on $k$), a {\emph{resonant
memory}} of sorts. If present in EMRIs, this resonant memory could have a large
impact on the gravitational wave phase as the system traverses a resonance.

\subsubsection{Matched Asymptotic Expansion}
\label{matched-asy}

The constant $b$ is fixed by the initial condition imposed at $x=0$, which
requires that a solution be valid in the $x\ll1$ limit. Recall that the solution
found in \eqref{first-order-y} is valid in the $x\gg1$ limit and that it
diverges as $x\to0$. Let us now look for a solution valid for $kx\ll1$, with $k
\ll1$, by using the ansatz
\be
\bar{y}(x)=\sum_{n=0}^\infty k^n\bar{y}_n(x)\,,\qquad{\rm{as}}\quad kx\ll1,k\ll 1\,,
\ee
where $\bar{y}_n(x)$ are undetermined functions independent of k. We could have
expanded $\bar{y}$ in $k(kx)^n$ instead of in $k^n$, but this would lead to more
complicated differential equations, although the solutions would be the same.
Here, we choose initial conditions $y_n(0)=0$ for all $n$, but the extension to
more general initial conditions is trivial. 

The zeroth-order solution ($n=0$) satisfies
\be
\bar{y}'_0(x)=1\qquad\Rightarrow\qquad\bar{y}_0(x)=x\,.
\ee
The first-order solution ($n=1$) satisfies
\be
\bar{y}_1'(x)=\cos(x^2)\qquad\Rightarrow\qquad\bar{y}_1(x)=\sqrt{\frac{\pi}{2}}
C\left(\sqrt{\frac{2}{\pi}} \, x\right)\,.
\ee
To next order in $k$, we substitute the solutions found thus far into the
differential equation
\be
\cos x^2+k \bar{y}_2' =\cos\left(x^2+k x \bar{y}_1(x)+k^2 x \bar{y}_2(x)\right).
\ee
We can expand the cosine using the assumption $kx\ll1$ to find that
\be
\bar{y}_2'(x)=-x\sin(x^2)\bar{y}_1\,, 
\ee
which then leads to the solution
\be
\bar{y}_2(x)=\frac{1}{2}\sqrt{\frac{{\pi}}{2}}\cos(x^2)C\left(\frac{x}{\sqrt{\pi
}}\right)-\frac{\sqrt{\pi}}{8}C\left(\frac{2}{\sqrt{\pi}}x\right)-\frac{x}{4}.
\ee
Similarly, the equation satisfied by the third-order term in the expansion is
\be
\bar{y}_3'(x)=-\frac{x^2}{2}\cos(x^2)\bar{y}_1^2(x)-x\sin(x^2)\bar{y}_2(x)\,,
\ee
but this cannot be explicitly integrated. Putting together all the pieces found
so far, we get
\begin{align}
\bar{y} &= x+k\sqrt{\frac{\pi}{2}}C\left(\sqrt{\frac{2}{\pi}}x\right)\nonumber\\
&+ k^{2}\left[\frac{1}{2}\sqrt{\frac{\pi}{2}}\cos(x^{2}) C(\frac{x}{\sqrt{\pi}})
-\frac{\sqrt{\pi}}{8}C\left(\frac{2 x}{\sqrt{\pi}}\right)-\frac{x}{4} \right]\,.
\label{first-order-y-inner}
\end{align}

Let us now asymptotically match this solution to the one found in
\eqref{first-order-y}. For such a procedure to be valid, a buffer zone must
exist where both solutions are simultaneously valid. Since \eqref{first-order-y}
was found by assuming $x\gg1$, while \eqref{first-order-y-inner} assumes that
$kx\ll 1$, this implies that a buffer zone does exist with extension $1\ll x\ll
k^{-1}$. Asymptotic matching requires that we asymptotically expand
\eqref{first-order-y} in $k x \ll 1$ and \eqref{first-order-y-inner} in $x\gg1$,
and then set these two expansions equal order by order. To leading order, we
find that
\be
x+\frac{\sqrt{2 \pi}}{4} k \sim x + b\,,
\ee
with remainders of relative ${\cal{O}}(1/x,kx)$. This immediately leads to 
\be
b \sim \frac{\sqrt{2 \pi}}{4} k.
\ee

\subsection{Phase Resonances}
\label{phase-res-lead-order}

Let us first consider the solution to the phase resonance equation, \eqref{diffeq2}, in the $k>1$ case. A constant
solution as $x\gg1$ exists provided $\phi''=\phi'=0$, which is satisfied if
\ba
\phi &\sim& \arccos(-1/k)\,, \nonumber \\
\phi'(0) &\sim& \arccos(1/k)- \sqrt{k^{2} - 1} - \pi\,.
\ea 
The first condition is the same one that $xy$ had to satisfy in the frequency resonance
case, but the second condition now imposes a constraint on the initial
conditions. If one chooses the initial conditions $\phi(0)=\phi'(0)=0$, the
second constraint leads to
\be
\sqrt{k^2-1}=2n\pi-\arccos(-1/k)\qquad\mbox{for}\;\;n\in\mathbb{Z}^+
\label{algebraic-equation}
\ee
where $\arccos(x)$ is the principal value of the inverse cosine, taking values
in the range $[0,\pi]$. Expanding for $k\gg1$ we find the approximate solution
\be
k_{\rm th}=-\frac{\pi}{4}\left(1-n\pi\right)+\frac{1}{4}\left[\pi^{2}-8n\pi^2
+16 n^2 \pi^2-8\right]\,.
\label{kthreshold}
\ee
Evaluating this expression for the first few values of $n=1,2,\ldots$, we find
that $k_{\rm th}\sim4.60378,\,10.9499,\,\ldots$, which are to be compared with
the exact numerical solutions to \eqref{algebraic-equation}, which are $k_{\rm th}=
4.60334,\,10.9499,\,\ldots$. We see that the error in the above asymptotic
expansion goes roughly as $1/k^5$.

Let us now consider the solution to \eqref{diffeq2} in the $k<1$ case. As
before, we concentrate on perturbative solutions in $k\ll1$. The zeroth-order
solution is found by setting $k=0$ in \eqref{diffeq2}: $\phi\sim\phi_0\equiv x^2
/2$. The first-order solution in $k$ can be found by postulating that
\be
\phi\sim\phi_0(x)+\phi_1(x)\,.
\ee
Inserting this into \eqref{diffeq2} we have
\be
\phi_{1}''=k\cos\left( \frac{1}{2} x^{2} + \phi_{1} \right)\,.
\ee
We assume that $\phi_1$ is subdominant relative to $\phi_{0}$, and so we
approximate the argument of the cosine as $x^2/2$. We can then solve exactly for
$\phi_{1}$ to find that
\be
\phi_{1} = k \left[ x \sqrt{\pi} C\left(\frac{x}{\sqrt{\pi}}\right) - \sin\left(\frac{x^{2}}{2} \right) \right]\,,
\ee
where $C(x)$ is the Fresnel cosine function defined in~(\ref{fresCdef}). 

We can now compare this solution to the one obtained for frequency resonances.
Differentiating and taking the $x\gg 1$ limit, we find that
\be
\phi' \sim x+\frac{\sqrt{\pi} k}{2} + {\cal{O}}(x^{-1})\,.
\ee
Notice that as $x\to\infty$ this agrees with \eqref{first-order-y} in functional
form, but not in slope. That is, as the system goes through a phase resonance,
the slope is not corrected by $k$. We will find in Sec.~\ref{higher-order} that
this remains true as one calculates the solution to higher order in $k$.
Therefore, although frequency resonances seem to induce a memory, phase
resonances do not. 

\section{Higher-Order Asymptotic Behavior for $x \gg 1$ and $k \ll 1$}
\label{higher-order}

In this section we consider higher-order solutions in the $x\gg 1$ limit for
both phase and frequency resonances, and then compare these to numerical
solutions.

\subsection{Frequency Resonances}

In order to obtain a higher-order solution, we must construct an ansatz that
eliminates the $x$-dependent part of the right side of
\eqref{preliminary-step3}. We thus pose the ansatz $y\sim y_{\rm asy}=y_1+y_2$,
where $y_{1}$ is given in \eqref{first-order-y}, while $y_{2}$ is 
\be
y_2 \equiv \frac{a_{2}}{x} \sin\{2[(1+c)x^2+bx]\}\,.
\ee
Inserting this ansatz into \eqref{diffeq}, we find that
\ba
&& c + 2 (1+c) a_1 C_{1} + 4 (1 + c) a_{2}C_{2} + {\cal{O}}(x^{-1})\nonumber \\
&\sim& k \cos[(1 + c) x^{2} + bx] \cos(a_{1} S_{1} + a_{2} S_{2})\nonumber \\
&-& k \sin[(1 + c) x^{2} + bx] \sin(a_{1} S_{1} + a_{2} S_{2})\,,
\ea
where we use the notation
\begin{align}
S_{n} &\equiv \sin\{n[(1+c) x^2 +b x]\}\,, \nonumber \\
C_{n} &\equiv \cos\{n[(1+c)x^2 + b x]\}\,.
\end{align} 
As before, we note that $S_{n}$ and $C_{n}$ are bounded by unity, and since $a_1
={\cal{O}}(k)$ we expect that $a_{2} = {\cal{O}}(k^{2})$ or smaller. This
suggests that we can expand the cosine on the right side of the above equation
as in Sec.~\ref{lead-order-sec} to obtain
\be
c+(1+c)\left(2a_1C_1+4a_2C_2\right)\sim-\frac{a_1k}{2}\left(1+C_2\right)+kC_1\,,
\ee
to ${\cal{O}}(x^{-1},k^3)$. We see then that our previous solution still holds:
$c=-ka_1^2/2$ and $2(1+c)a_1=k$, implying that $c\sim-k^2/4$. We also see that
$4(1+c)a_2=a_1k/2$, which implies that $a_2=k^2/[16(1+c)^2]$ or simply that
$a_2\sim k^2/16$ when expanding in $k\ll1$. The second-order
solution therefore becomes 
\be
y_2\equiv\frac{k^2}{16x}\sin\left[2\left(1-\frac{k^2}{4}\right)x^2+2bx\right]\,.
\ee

We can obtain the next-order solution by constructing the ansatz $y\sim y_{\rm
asy}=y_1+y_2+y_3$, where $y_3\equiv\frac{a_3}{x}S_{3}$ and where we assume that
$a_3={\cal{O}}(k^3)$. Inserting this into \eqref{diffeq}, we get
\ba
&&c+2(1+c)a_1C_1+4a_2(1+c)C_2+6a_3(1+c)C_3\nonumber \\
&\sim& k C_1\cos(a_1 S_1 +a_2 S_2 +a_3 S_3) \nonumber \\
&-& k S_{1} \sin(a_{1} S_{1} + a_{2} S_{2} + a_{3} S_{3})\,.
\ea
Expanding the above equations in $a_{1} \ll 1$ and $a_{3} \ll 1$, we find that
\ba
&&c+2(1+c)a_1C_1+4a_2(1+c)C_2+6a_3(1+c)C_3\nonumber \\
&\sim& -\frac{ka_1}{2}+\left(k-\frac{ka_2}{2}-\frac{ka_1^2}{8}\right)C_1
+\frac{k a_{1}}{2} C_{2}\nonumber \\
&+& \left(\frac{k a_{1}^{2}}{8} + \frac{k a_{2}}{2} \right) C_{3}\,.
\ea
Matching cosine coefficients, this leads to the following system of equations:
\ba
c &\sim& - \frac{k a_{1}}{2}\,, \nonumber \\
2a_1(1+c)&\sim& k\left(1-\frac{a_2}{2}-\frac{a_1^2}{8}\right)\,,\nonumber \\
4 a_{2} (1 + c ) &\sim& k \frac{a_{1}}{2}\,, \nonumber \\
6 a_3 (1+c) &\sim& k \left(\frac{a_{1}^{2}}{8} + \frac{a_{2}}{2} \right)\,,
\ea
which we can solve as an expansion in $k$ to find that
\ba
a_{1} &\sim& \frac{k}{2}\,, \qquad a_{2} \sim \frac{k^{2}}{16}\,, \nonumber \\
a_{3} &\sim& \frac{k^{3}}{96}\,, \qquad c\sim-\frac{k^2}{4}-\frac{3}{64} k^4\,.
\ea
Therefore, our solution to third order becomes
\ba
y_{\rm asy} &=& b + \left(1- \frac{k^{2}}{4} -\frac{3}{64}k^4 \right)x \nonumber \\
&+& \frac{k}{2} \frac{1}{x} \sin\left[\left(1- \frac{k^{2}}{4}-\frac{3}{64}
k^{4}\right)x^2 + bx \right]\nonumber \\
&+& \frac{k^{2}}{16}\frac{1}{x} \sin\left[2\left(1-\frac{k^{2}}{4}-\frac{3}{64}
k^{4}\right)x^2+ 2bx \right] \nonumber \\
&+& \frac{k^{3}}{96}\frac{1}{x}\sin\left[3 \left(1 -\frac{k^{2}}{4}-\frac{3}{64}
k^{4}\right) x^{2} + 3 b x \right]\,.
\label{full-asy-sol}
\ea
Notice that this higher-order solution retains the resonant memory computed in
the previous section (that is, the $k^2$ correction to the linear-in-$x$ term,
which is dominant at late times). 

This procedure can be generalized to arbitrary high order in $k\ll1$. To leading
order in $1/x$, we make the ansatz $y\sim y_{(1)}$, where
\be
y_{(1)}\equiv(1+c)x+b+\frac{k}{x} \sum_{n=1}^{\infty} a_n^{(1)} S_n + b_n^{(1)}
C_n + O\left(\frac{1}{x^2}\right)\,,
\label{hi-order-ansatz}
\ee
where $(a_n^{(1)},b_n^{(1)},b,c)$ are constant coefficients that depend on $k$.
We can expand $(a_n^{(1)},b_n^{(1)})$ as an expansion in $k$, that is,
$a_n^{(1)}=\sum_m a_{nm}^{(1)} k^m$, and solve for these coefficients by
equating the coefficients of the $S_n$'s and $C_n$'s at different orders in $k$.
Because the derivative of \eqref{hi-order-ansatz} is 
\be
y'_{(1)} = 1 + c + \sum_{n=1}^{\infty} \frac{2 n (1 + c) k}{x}\left(a_n^{(1)}
C_n - b_n^{(1)} S_n\right) + O\left(\frac{1}{x^2}\right)\,.
\ee
we can evaluate \eqref{diffeq} to obtain
\ba
y'_{(1)} &=& 1 + k \; C_{1}\cos\left[k \sum_{n=1}^{\infty} a_n^{(1)} S_n +
b_n^{(1)} C_n\right] \nonumber \\
&-& k \; S_{1}\sin\left[k\sum_{n=1}^\infty a_n^{(1)}S_n+b_n^{(1)}C_n\right]\,.
\ea
The generic sine and cosine Taylor expansion formula allows us to rewrite the
above equation as
\ba
y'_{(1)} &=& 1+k\;C_1\sum_{\ell=0}^\infty \frac{(-1)^{\ell}k^{2\ell}}{(2 \ell)!}
\left[\sum_{n=1}^\infty a_n^{(1)}S_n+b_n^{(1)} C_n\right]^{2\ell}\nonumber \\
&-& k \; S_{1}\sum_{\ell=0}^\infty \frac{(-1)^{\ell+1}k^{2\ell+1}}{(2\ell+1)!} 
\left[\sum_{n=1}^{\infty} a_n^{(1)} S_n + b_n^{(1)} C_n\right]^{2\ell + 1}\,.
\ea
At this point, no further progress can be achieved because one needs to evaluate
the $(2\ell+1)$st and the $(2\ell)$th power of an infinite series, which is not
easy to do in closed form. This is why it is more convenient to expand the first
few terms in the series, as done earlier in this section.
 
We can generalize the previous procedure to higher order in $x$.
Postulate the ansatz $y \sim y_{(1)} + y_{(2)}$, where 
\be
y_{(2)} \sim \frac{k}{x^2} \sum_{n=0}^\infty a_n^{(2)} S_n + b_n^{(2)} C_n .
\ee
The derivative of $y$ is then simply
\ba
y' &=& 1+c+\sum_{n=1}^\infty \frac{2n(1+c)k}{x}\left(a_n^{(1)}C_n-b_n^{(1)}
S_n\right) \\ \nonumber
&-& k \sum_{n=1}^{\infty}\left[\frac{2 n (1 + c) b_{n}^{(1)} + a_{n}^{(1)}}{x^{2}}\right]S_{n} 
\\ \nonumber \nonumber
&+& k \sum_{n=1}^{\infty} \left[\frac{2 n (1 + c) a_{n}^{(1)} - b_{n}^{(1)}}{x^{2}}\right]C_{n} 
+ O\left(\frac{1}{x^3}\right)\,,
\ea
while the right side of \eqref{diffeq} implies that
\ba
y' &=&1+k \; C_1\cos\left[k\sum_{n=1}^\infty \left(a_n^{(1)}S_n+b_n^{(1)}
C_n\right)\right]\nonumber \\
&-& k \; S_{1}\sin\left[k \sum_{n=1}^\infty \left(a_n^{(1)}S_n+b_n^{(1)}C_n
\right)\right] \\ \nonumber 
&-& \frac{k^{2}}{x} C_{1}\sin\left[k \sum_{n=1}^{\infty} \left(a_n^{(1)} S_n
+ b_n^{(1)} C_n\right)\right] \\ \nonumber 
&\times& \sum_{n=1}^{\infty} \left(a_n^{(2)} S_n + b_n^{(2)} C_n\right) 
\nonumber \\ \nonumber &+& \frac{k^{2}}{x} S_{1}
\cos\left[k\sum_{n=1}^\infty \left(a_n^{(1)} S_n + b_n^{(1)} C_n \right)\right]
\\ \nonumber 
&\times& \sum_{n=1}^{\infty} \left(a_n^{(2)} S_n + b_n^{(2)} C_n\right)\,.
\ea
One could now expand these equations in $k \ll 1$ and equate coefficients to get
equations for the $a_n^{(2)}$'s and $b_n^{(2)}$'s. Following this scheme, one
can find the subdominant terms in the asymptotic expansion of the solution as
series in $1/x$. 

\subsection{Phase Resonances}

Let us now concentrate on higher-order solutions to the phase-resonance equation
in the $k<1$ case. We thus postulate the ansatz $\phi\sim\phi_{\rm asy}$, where
\be
\phi_{\rm asy} = \phi_{0}(x) + \phi_{1}(x) + \phi_{2}(x)\,,
\label{AsySolPhase}
\ee
where we recall that $\phi_0=x^2/2$, and we rewrite $\phi_1$ as
\be
\phi_1=k\frac{\sqrt{\pi}}{2}x+k\left\{\sqrt{\pi}x\left[C\left(\frac{x}{
\sqrt{\pi}}\right)-\frac{1}{2}\right]-\sin\left(\frac{x^2}{2}\right)\right\}\,.
\label{psi1sol}
\ee
We have factored out the unbounded-in-$x$ part of $\phi_1$ and the second term
is now bounded for all $x$ and tends to $0$ as $x\to\infty$. This ensures that the term in curly brackets is small for
sufficiently small $k$ as $x\to\infty$. We can then see that $\phi_2$ must
satisfy the differential equation
\be
\phi_2''\sim k\cos\left(\phi_0+\phi_{1} \right)\,.
\ee
where we seek solutions accurate to $O(k^2)$ and, as in
Sec.~\ref{phase-res-lead-order}, we neglect the $\phi_2$ term in the source.
Inserting $\phi_0$ and $\phi_1$ and using the fact that the bracketed term in
(\ref{psi1sol}) is everywhere small to expand the cosine, we find that
\begin{align}
\phi_2'' &= k\cos\left(\frac{x^2}{2} + k\frac{\sqrt{\pi}}{2}x\right)\nonumber \\
&- k^2 \sin\left(\frac{x^2}{2} + k\frac{\sqrt{\pi}}{2}x\right)\nonumber \\ 
&\times \left\{\sqrt{\pi} x \left[ C\left(\frac{x}{\sqrt{\pi}}\right) -\frac{1}
{2}\right] - \sin\left(\frac{x^2}{2}\right)\right\}.
\end{align}
Integrating this equation twice, imposing the condition that $\phi_2(0)=\phi_2'
(0)=0$, and ignoring terms explicitly proportional to $k^3$ or higher, we obtain
\begin{align}
\phi_{\rm asy} &= \frac{x^{2}}{2} 
+ k\sqrt{\pi} x \left[\frac{1}{2} - C\left(\frac{k}{2}\right) + k\frac{(\sqrt{2}-1)}{2\sqrt{2}}\right] 
+\frac{\pi}{8}k^2 \nonumber \\
&+ k\left\{\sqrt{\pi}\left(x+\frac{\sqrt{\pi} k}{2}\right) \left[C\left(\frac{x}
{\sqrt{\pi}}+\frac{k}{2}\right)-\frac{1}{2}\right] \right.\nonumber \\ 
&-\left.\sin\left[\frac{1}{2}\left(x+k\frac{\sqrt{\pi}}{2}\right)^2\right]
\right\}\nonumber \\
&+k^2\left\{-x \sqrt{\frac{\pi}{2}} \left[C\left(\frac{\sqrt{2} x}{\sqrt{\pi}}
\right) - \frac{1}{2}\right] + \frac{1}{2}\sin(x^2) \right.\nonumber \\
& \left. +\frac{\pi}{2} \left[C\left(\frac{x}{\sqrt{\pi}}\right)\right]^2 -
\frac{\pi}{2} C\left(\frac{x}{\sqrt{\pi}}\right) +\frac{\pi}{8} \right\}\,.
\label{psi2}
\end{align}
The slope of this solution for large $x$ gives the $k$-correction to the
gradient
\be
\phi_{\rm asy}' \sim x + \frac{\sqrt{\pi}}{2} k \left[1-\frac{k}{\sqrt{2}} +
{\cal{O}}(k^{2}) \right].
\ee
Notice that although the constant is $k$-corrected, the linear-in-$x$ term is
not, showing again that phase resonances do not acquire a resonant memory
imprint. 

We proceed to higher order, and by analogy we write down the solution for
$\phi_2(x)$ as the sum of the part on the first line of (\ref{psi2}) that grows
linearly with $x$ and a part that is bounded for all $x$ and has a convergent
integral on the range $[0,\infty]$. Schematically, this takes the form
\be
\phi_{1} + \phi_2=-\frac{\pi}{8}k^2+\left(\frac{\sqrt{\pi}}{2}k -
\frac{\sqrt{\pi}}{2\sqrt{2}} k^2\right) x + k \Phi_1(x) + k^2\Phi_2(x),
\ee
where $k\Phi_1(x)$ is the term on the second and third lines of \eqref{psi2},
while $k^2\Phi_2(x)$ is the term on the fourth and fifth lines. Notice that $k
\Phi_1(x)$ does contain terms of $O(k^2)$. We now seek the next order solution,
$\phi_3$, that solves the equation
\be
\phi_3''=k\cos\left(\frac{1}{2}x^2 + \phi_1 + \phi_2\right)\,,
\ee
where again we have neglected $\phi_{3}$ in the source term. Inserting the
solution known so far and expanding the cosine, keeping terms up to $O(k^3)$, we
find that
\begin{align}
\phi_3''&=k\cos\left[\frac{1}{2} x^2 + \left(\frac{\sqrt{\pi}}{2}k -
\frac{\sqrt{\pi}}{2\sqrt{2}} k^2\right) x+\frac{\pi}{8}k^2\right] \nonumber\\
&-\frac{k^3}{2}\cos\left(\frac{1}{2}x^2\right) \Phi_1^2 \nonumber \\
&-k^2\sin\left(\frac{1}{2}x^2+\frac{\sqrt{\pi}}{2} k x\right) \Phi_1\nonumber\\
&-k^3 \sin\left(\frac{1}{2} x^2\right) \Phi_2\,.
\label{d2psi3}
\end{align}
Note that in each term we have eliminated terms in the arguments of the cosine
and sine that are lower order than $k^{3-n}$, where $n$ is the order of the $k$
prefactor to the term. By integrating these terms over the range $[0,\infty]$,
we can derive the $O(k^3)$ correction to the asymptotic gradient. We find the
contributions to the $O(k^3)$ term in the gradient from each line of
(\ref{d2psi3}) are $\sqrt{\pi}/2\sqrt{2}$, $-0.06202$, $0$, and $\sqrt{\pi}(1-2
\sqrt{2}+\sqrt{3})/8-0.07844$ respectively. The final form for the asymptotic
gradient is then
\be
\phi_{\rm asy}'\sim\frac{\sqrt{\pi}}{2}k-\frac{\sqrt{\pi}}{2\sqrt{2}}k^2+
0.46484 k^3 \cdots .
\ee

\subsection{Comparison to Numerical Results}

\begin{figure*}
\includegraphics[height=8.5cm,clip=true,angle=-90]{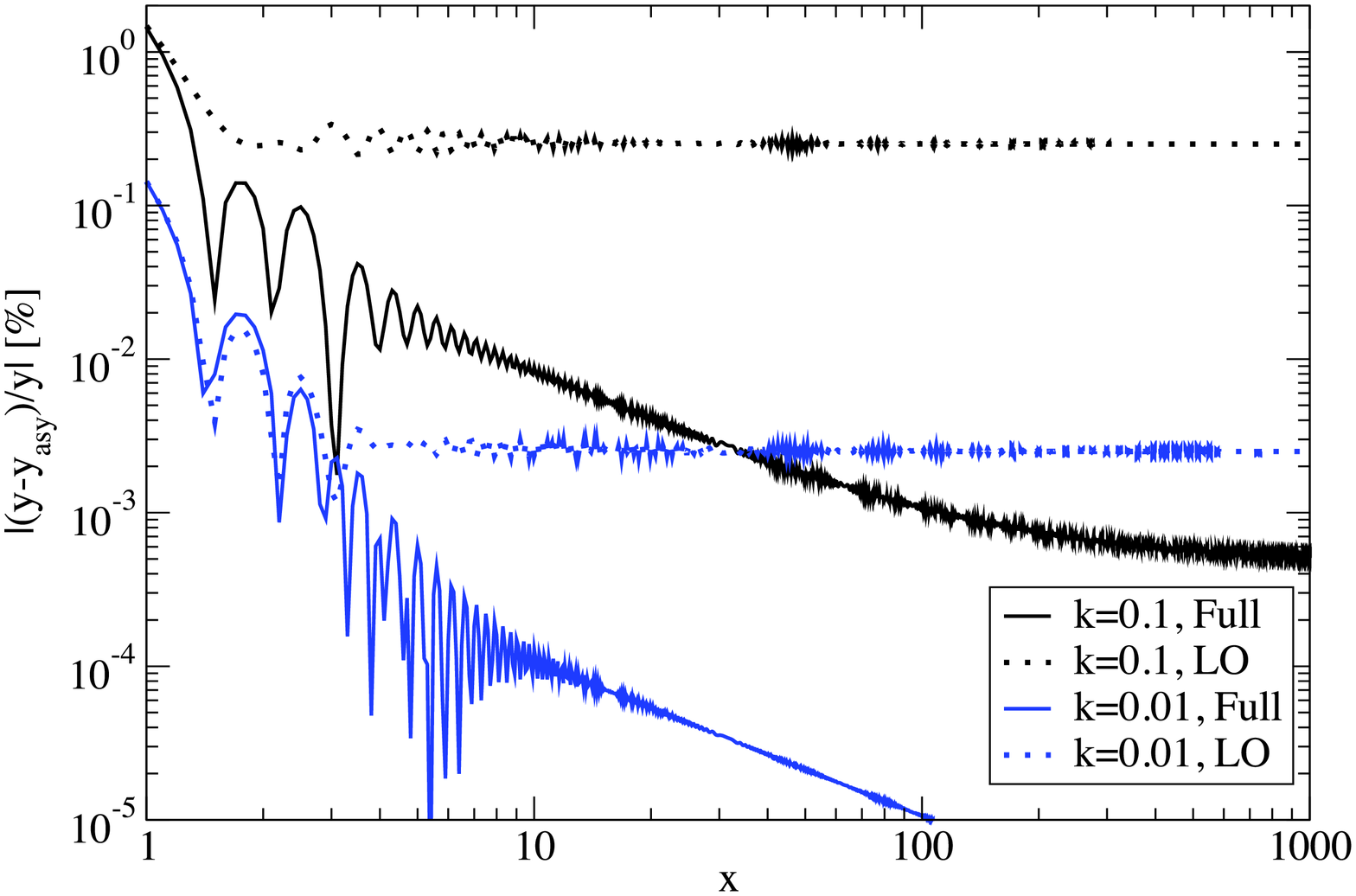}
\includegraphics[height=8.5cm,clip=true,angle=-90]{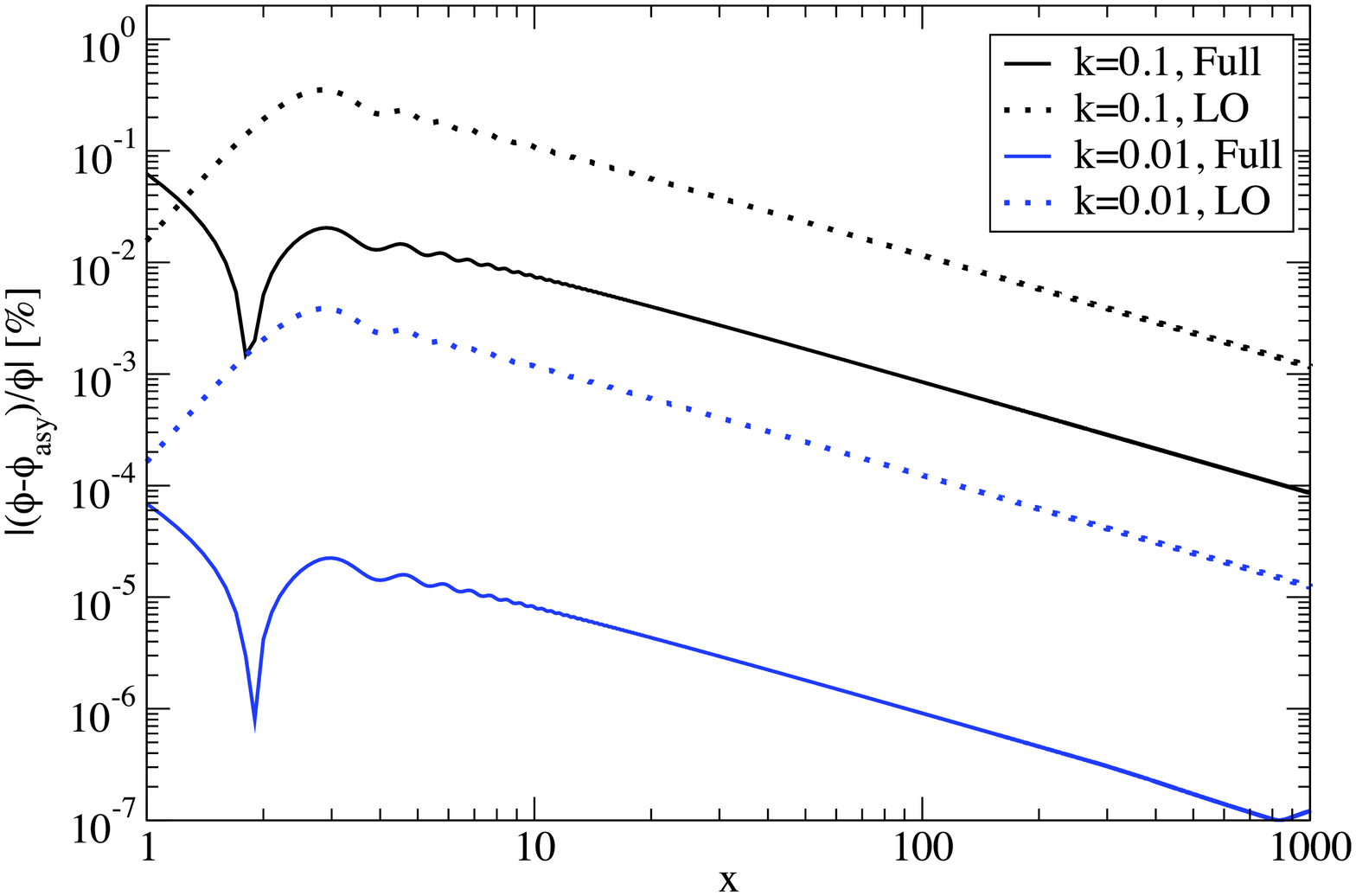}
\caption{\label{fig:Diffs} Left: Difference between the numerical solution,
$y$, to the frequency resonance equation and the asymptotic expansion in \eqref{full-asy-sol} to all computed orders
in $k$ (solid line) and to linear order in $k$ (dotted lines) as a function of
$x$. Right: Same difference as left-panel but for the solution, $\phi$, to the phase resonance equation. Observe that in both cases the full solution does much better than the ${\cal{O}}(k)$
truncation.}
\end{figure*}

The left panel of Fig.~\ref{fig:Diffs} shows the difference between the
numerical solution for $y$ and the asymptotic expansion in \eqref{full-asy-sol}
to all computed orders in $k$ (solid line) and to linear order in $k$ (dotted
lines). In all plots we have chosen $b$ to be that found in
Sec.~\ref{matched-asy}. As expected, the solid lines are much closer to zero
than the dotted ones. Moreover, notice that the asymptotic solutions found in
the limit $x\gg1$ are already quite good (better than $1\%$ relative to the
numerical solution) below $x<10$.

The right-panel of Fig.~\ref{fig:Diffs} shows the same type of difference as
the left-panel, but for the phase solution. As before, observe that the full
solution agrees with the numerical solution much better than its truncated
version. Unlike the $y_{\rm asy}$ solution, the $\phi_{\rm asy}$ is globally
valid, as we did not restrict attention to the $x \gg 1$ limit, and this can be clearly seen in Fig.~\ref{fig:Diffs}. It is also apparent that, as expected, in both cases the smaller $k$ is, the better the asymptotic solution.

\begin{figure}
\includegraphics[height=8.5cm,clip=true,angle=-90]{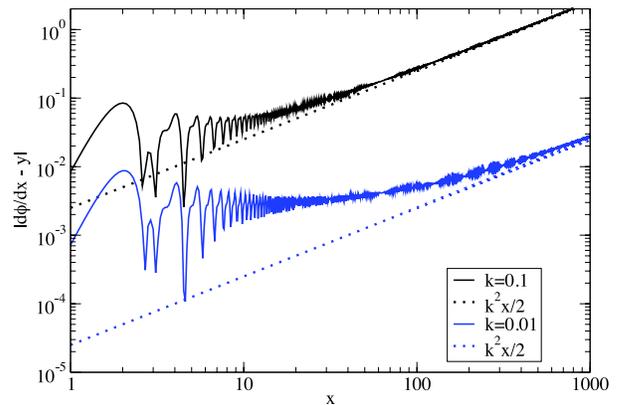}
\caption{\label{fig:NumDiff} Difference between the numerical solution for $y$
and the $x$-derivative of the numerical solution for $\phi$. For comparison, we
also plot the asymptotic slopes $k^{2} x/2$.}
\end{figure}

Although the previous figures establish that the asymptotic solutions are indeed
accurate representations of the numerical ones, they do not compare the $y$ and
$\phi$ solutions to each other. Figure~\ref{fig:NumDiff} shows the difference
between $y$ and $d\phi/dx$ computed numerically in both cases. As predicted by
the asymptotic solutions, the difference is approximately $k^{2} x/2$ (the
dotted lines) in the $x \gg 1$ limit.

\section{Asymptotic transition at $k=1$}
\label{transition}

As described in the previous sections, both the $y$ and $\phi$ solutions
experience a transition as $k \to 1$. In this section we discuss this
transition in more detail using hyper-asymptotic tools. 

\subsection{Frequency Resonances}

Let us try to understand the fundamental change in the asymptotic behavior of
the $y$ solution as $k$ transitions from $k>1$ to $k<1$ from above. To do so, we
must find the most important term as $k$ approaches $1$ to {\it all} orders in
powers of $1/x$. Then, we must sum the series and identify the singularity at
$k=1$.

Assuming that $k>1$, we know that the behavior of $y(x)$ in (\ref{diffeq}) as
$x\to\infty$ is described by a series in inverse odd powers of $x$:
\begin{equation}
y\sim\frac{a}{x}+\frac{b}{x^3}+\frac{c}{x^5}+\frac{d}{x^7}+\cdots
\qquad x\to\infty,
\label{e2}
\end{equation}
where 
\begin{equation}
\cos(a)=-\frac{1}{k}.
\label{e3}
\end{equation}
As discussed earlier, as $k\to1^+$, $a\to(2n+1)\pi$, and thus $\sin a$
approaches $0$. To higher order in $1/x$, one easily finds that
\ba
b&=&\frac{a}{k\sin(a)}\,, \label{e4}
\\ c&=&\frac{6a\sin(a)k+a^2}{2k^3[\sin(a)]^3}\,. \label{e11}
\ea
Thus, when $\sin(a)$ is small, the most singular part of $c$ is 
\begin{equation}
c\sim\frac{a^2}{2k^3[\sin(a)]^3}.
\label{e12}
\end{equation}
Similarly, the most singular part of $d,\,e,\,f,\,\ldots$ is
\begin{eqnarray}
d &\sim& \frac{a^3}{2k^5[\sin(a)]^5},\qquad
e \sim \frac{5a^4}{8k^7[\sin(a)]^7},\nonumber\\
f &\sim& \frac{7a^5}{8k^9[\sin(a)]^9},\qquad
g \sim \frac{21a^6}{16k^{11}[\sin(a)]^{11}},\nonumber\\
h &\sim& \frac{33a^7}{16k^{13}[\sin(a)]^{13}}\,.
\label{e13}
\end{eqnarray}

The numerical coefficients, $1$, $\frac{1}{2}$, $\frac{1}{2}$, $\frac{5}{8}$, $\ldots$, are given by a very simple formula:
\begin{equation}
F(n)=\frac{(2n)!}{n!(n+1)!2^n}\qquad(n=0,\,1,\,2,\,3,\,...).
\label{e14}
\end{equation}
Therefore, if we sum the most singular terms as $k\to1^{+}$ to all orders in
powers of $1/x$, we find that
\begin{equation}
y(x)\sim\frac{a}{x}+\frac{a}{kx^3\sin(a)}
\sum_{n=0}^\infty F(n)\left[\frac{a}{x^2k^2[\sin(a)]^2}\right]^n.
\label{e15}
\end{equation}
{}From the Taylor expansion
\begin{equation}
\frac{1}{z}-\frac{\sqrt{1-2z}}{z}=1+\frac{1}{2}z+\frac{1}{2}z^2+\frac{5}{8}z^3
+\ldots=\sum_{n=0}^\infty F(n)z^n
\label{e16}
\end{equation}
one finds that
\begin{equation}
y(x)\sim\frac{1}{x}\left[a+k\sin(a)-k\sin(a)\sqrt{1-\frac{2a}{x^2(k^2-1)}}
\,\,\right],
\label{e17}
\end{equation}
where we have used the identity $k^2[\sin(a)]^2=k^2-1$. This shows that there
is a square-root-branch-cut singularity where the asymptotic behavior goes
complex as $k\to1^{+}$.

In addition to this branch-cut singularity, one can also show that the
higher-order terms in $1/x$ bunch up into families as $x\to\infty$, with each
pair of families separated by an unstable separatrix curve. As shown earlier, as
$k\to1^+$, the leading-order slope of the solution $a\to(2n+1)\pi$ and $\sin a$
approaches $0$. There are, however, many solutions for $a$ in (\ref{e3}) as $k
\to 1^{+}$: the first lies just below $\pi$ (but above $\pi/2$); the second lies
just above $\pi$; the third and fourth lie just below and just above $3\pi$,
and so on. 

Consider {\emph {two different solutions}}, $y_1(x)$ and $y_2(x)$ corresponding
to {\emph {one}} of the infinite number of possible values of $a$ and define
\begin{equation}
Y(x)\equiv y_1(x)-y_2(x).
\label{e5}
\end{equation}
Observe that $Y(x)$ satisfies the differential equation
\begin{equation}
Y'(x)=k\cos\left[xy_1(x)\right]-k\cos\left[xy_2(x)\right].
\label{e6}
\end{equation}
Using the identity
\begin{equation}
\cos\alpha-\cos\beta=-2\sin\left(\frac{\alpha+\beta}{2}\right)
\sin\left(\frac{\alpha-\beta}{2}\right),
\label{e7}
\end{equation}
we can rewrite (\ref{e6}) as
\begin{equation}
Y'(x)\sim-2k(\sin a)\sin\left[\frac{xY(x)}{2}\right],
\label{e8}
\end{equation}
for large $x$.

Let us now make the assumption that $y_1(x)$ and $y_2(x)$ approach one another
as $x$ gets large, so that $Y(x)$ is small when $x>>1$. Then, this differential
equation becomes
\begin{equation}
Y'(x)\sim x(\tan a)\,Y(x),
\label{e9}
\end{equation}
whose solution is 
\begin{equation}
Y(x)\sim C \; e^{\frac{1}{2} x^{2} \tan a }.
\label{e10}
\end{equation}
Note that this solution is {\it growing} exponentially if $\tan a$ is positive, 
and thus the assumption that $Y(x)$ is small as $x\to\infty$ is not valid.
This is the unstable (separatrix) case. However, if $\tan a$ is negative, then 
we have the stable case, and we have shown that the family of solutions
corresponding to this case all bunch together {\emph{exponentially fast}}. 
Note that there is an alternation between stable and unstable behavior:
Stable behavior occurs only for the values of $a$ that are just {\emph{below}}
$(2n+1)\pi$ for $n \in {\mathbb{Z}}$, that is, $\pi$, $3\pi$, $5\pi$, and so on,
while unstable behavior occurs for the values of $a$ just above $(2n+1)\pi$.

\subsection{Phase Resonances}

\begin{figure*}[ht]
\includegraphics[width=6cm,clip=true,angle=-90]{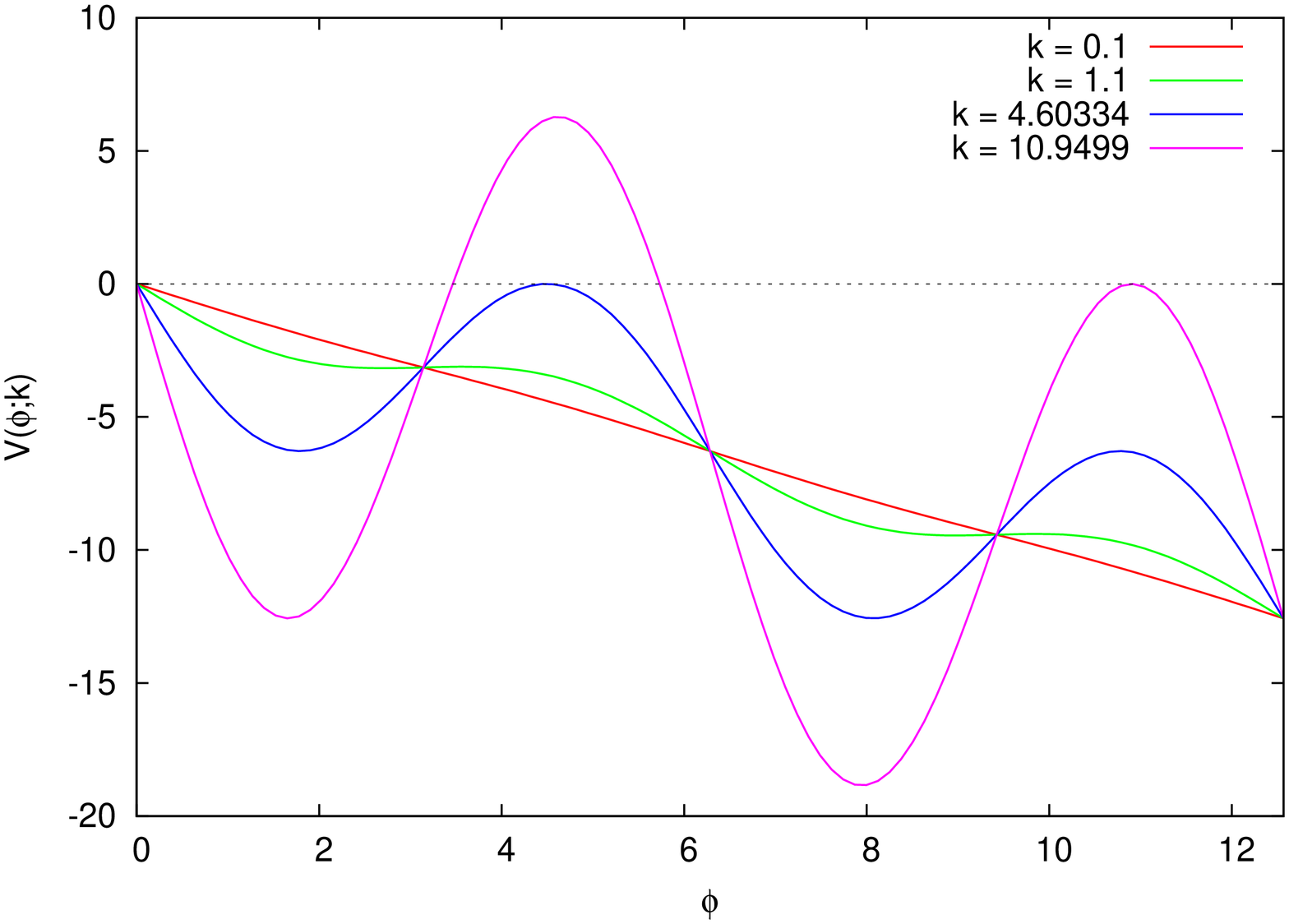} 
\includegraphics[width=6cm,clip=true,angle=-90]{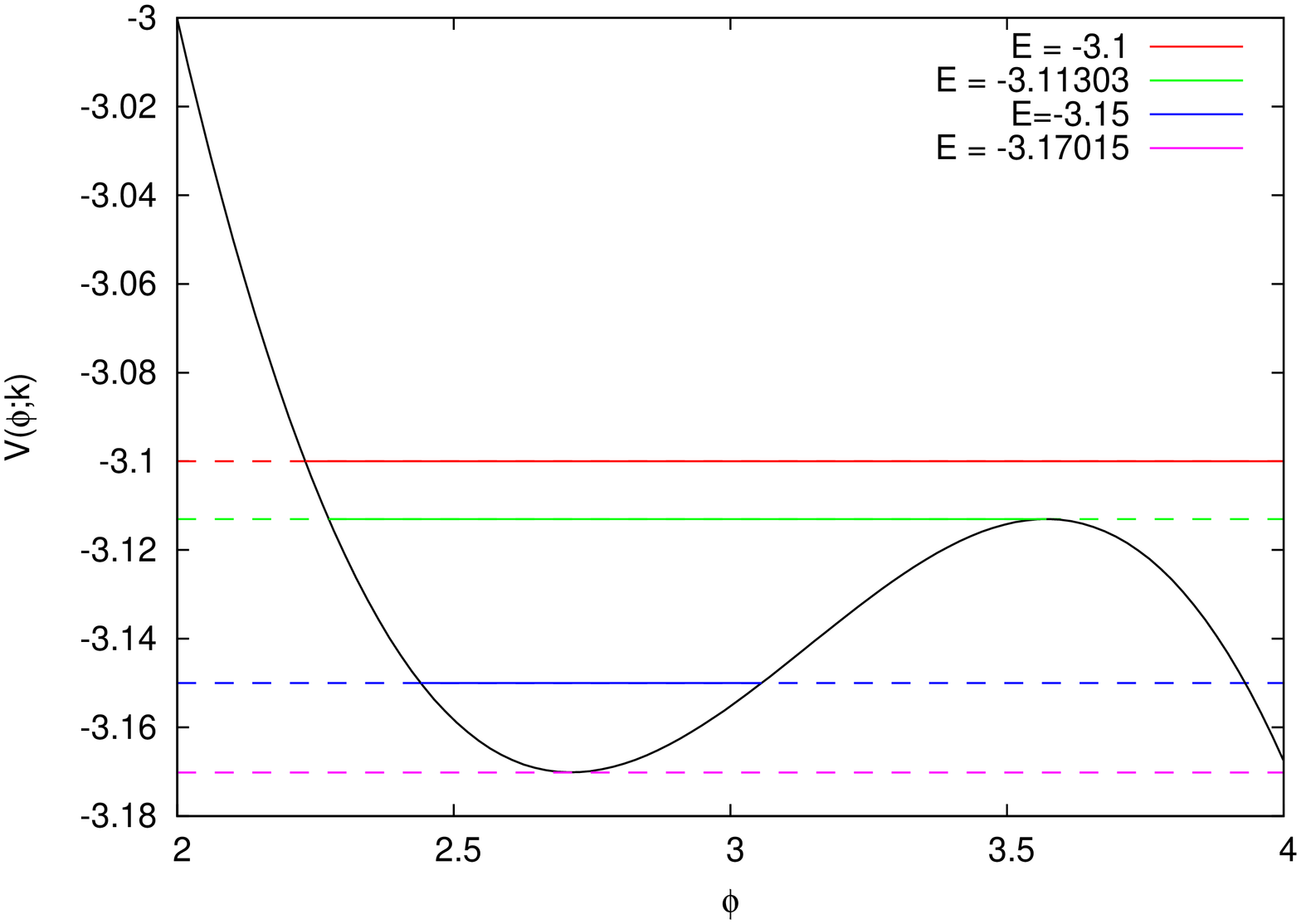}
\caption{\label{phieffpot}Effective potential for the phase resonance
equation, as defined by Eq.~(\ref{effpotdef}). The left panel shows the
potential for four different values of $k$, while the right panel shows a
close up of the region $2 \leq \phi \leq 4$ when $k=1.1$. Each horizontal line
corresponds to a particular choice of the constant $E$. Motion can only exist
where $E > V(\phi;k)$, as indicated by the solid parts of the lines shown.}
\end{figure*}

The phase solutions also show a fundamental qualitative change of behavior as
$k$ transitions from $k>1$ to $k<1$. Because the phase differential equation,
\eqref{diffeq2}, admits a first integral of the motion, the analysis is simpler
than in the frequency case and it does not require a hyper-asymptotic analysis.
This is best understood if we consider the first-order form of \eqref{diffeq2},
which we rewrite here as
\be
\frac{1}{2}(\phi')^2 = E - V(\phi ; k)\,,
\label{effpotdef}
\ee
where $V(\phi ; k) = -\phi - k \sin\phi$ and $E = \phi'(0)$. The potential is
shown for representative values of $k$ in the left panel of
Figure~\ref{phieffpot}. Motion can only exist in regions where $E>V(\phi)$. The
initial conditions $\phi(0)=\phi'(0)=0$ correspond to $E=0$. For $k<1$, the
effective potential has no turning points and for any initial conditions the
motion will be unbounded with $\phi\rightarrow+\infty$. When $k>1$, the
potential does have turning points. The right panel of Figure~\ref{phieffpot}
shows a close-up of the potential for $k=1.1$ in the vicinity of the turning
points. For $E > -3.11303$, the motion is unbounded as in the $k < 1$ case.
However, for $-3.11303>E>-3.17015$, the motion intersects the potential twice,
and it can be oscillatory for suitable initial values of $\phi$, or unbounded if
$\phi(0)$ is sufficiently large. For $E < -3.17015$, the motion is again only
unbounded for sufficiently large $\phi(0)$.

These specifications for $E$ and $\phi(0)$ place restrictions on the initial
conditions, which are inconsistent with the conditions we want to impose. For
the conditions $\phi(0)=\phi'(0)=0$, the motion is unbounded for $k\lesssim
4.60334$, the critical value computed in \eqref{kthreshold}. When $k\approx
4.60334$, the potential intersects the $E=0$ axis a second time. The motion will
asymptotically approach the limiting value $\phi_\infty\approx 4.49341$. For $k
\gtrsim 4.60334$, the motion is oscillatory. At the next limiting solution for
$k$, $k\approx10.9499$, the effective potential has another intersection with
the $E=0$ axis. However, this region is inaccessible to motion with these
initial conditions and the motion is still oscillatory. Starting the motion with
$\phi'(0)=0$ and $5.73224\lesssim\phi(0)\lesssim 10.9041$ would generate a
solution that asymptotically approaches the next limiting value $\phi_\infty
\approx10.9041$.

At the threshold value $k\approx4.60334$, two solutions $\phi_1(x)$, $\phi_2(x)$
with $0>E_1\neq E_2<0$ will oscillate with different frequencies, and we
therefore expect the difference $\phi_1(x)-\phi_2(x)$ to be oscillatory. If
$E_1<0<E_2$, one solution will be unbounded and therefore the difference will
grow like $x^2/2$. For $0<E_1\neq E_2>0$, both solutions are unbounded, and we
expect the difference to grow linearly with $x$. 

\section{Generalized Resonances}
\label{gen-res}

Section~\ref{intro} derived certain equations [Eqs.~\eqref{diffeq} and
\eqref{diffeq2}] that are representative of phase and frequency resonances in
EMRIs, but in doing so we made two important simplifying assumptions: (b) we
ignored the sine term in the sum given in Eq.~(\ref{probdef}) and (c) we ignored
higher-$\ell$ terms in this same sum. In this section we relax these two
assumptions and discuss how the solutions are modified.

Let us first relax assumption (b). If the sine term is included, \eqref{diffeq}
and \eqref{diffeq2} can be written as 
\be
\frac{{\rm d}y}{{\rm d}x} = 1 + k \cos(xy + \delta\phi), \qquad \frac{{\rm d}^2
\phi}{{\rm d}x^2} = 1 + k \cos(\phi + \delta \phi)\,,
\ee
where $\delta \phi$ is a constant. We can repeat the analysis of the frequency
evolution equation with the modification introduced above by making the ansatz 
\be
y_1 \equiv (1+c)x + b + \frac{a_1}{x} \sin[(1+c)x^2 + bx + \delta \phi]
\ee
in place of (\ref{freqLOansatz}). The analysis proceeds exactly as before, but
with the arguments of the various cosine and sine terms modified via $(1+c)x^2+
bx\to (1+c)x^2+bx+\delta\phi$. The asymptotic slope, $(1+c)$, is unchanged as a
function of $k$. The solution for $b$ will be modified, however, because the
expansion for $kx\ll1$ described in Sec.~\ref{matched-asy} is modified. In
particular
\begin{align}
\bar{y}_0(x) &=x\,, 
\nonumber \\
\bar{y}_1(x) &=\cos(\delta \phi) \sqrt{\frac{\pi}{2}} C\left(\sqrt{\frac{2}{\pi}
}x\right)-\sin(\delta\phi)\sqrt{\frac{\pi}{2}}S\left(\sqrt{\frac{2}{\pi}}x
\right)\nonumber \\
\bar{y}_2(x) &= \frac{1}{2} \bar{y}_1 - \frac{x}{4} - \frac{\sqrt{\pi}}{8}
\left[ \cos(\delta \phi) C\left( \frac{2}{\sqrt{\pi}}x\right) \right.\nonumber\\
& \left. - \sin(\delta \phi) S\left( \frac{2}{\sqrt{\pi}}x\right)\right]\,,
\end{align}
in which $S(\cdot)$ denotes the Fresnel sine function, defined by a similar equation to Eq.~(\ref{fresCdef}), but with the cosine replace by a sine. After asymptotic
matching, we then find that $b\to\sqrt{\pi}(k/2)\cos({\pi/4+\delta\phi})$.

In the phase-resonance case, the addition of the $\delta\phi$ to the equation of
motion is equivalent to solving the original problem with a modified initial
condition: $\phi(0)=\delta \phi$ and $\phi'(0)=0$. However, the solution to the
modified equation with the standard initial condition $\phi(0)=0$ can also be
found straightforwardly using the method described in this paper. In that case,
the zeroth-order-in-$k$ solution is unchanged, $\phi_0=x^2/2$, but the
first-order-in-$k$ correction, $\phi_1$, is modified to
\ba
\phi_1&=&k\sqrt{\pi} x \left[ \cos(\delta \phi) C\left(\frac{x}{\sqrt{\pi}}
\right)-\sin(\delta \phi) S\left(\frac{x}{\sqrt{\pi}}\right)\right]\nonumber \\
&& -k\sin\left(\frac{x^2}{2}+\delta \phi\right) + k\sin(\delta \phi)
\ea
from which we see that the asymptotic correction to the gradient is modified to
$k\sqrt{\pi/2}\cos(\delta \phi+\pi/4)$. Continuing to the next order in $k$,
we obtain for the asymptotic gradient
\begin{align}
\frac{{\rm d}\phi}{{\rm d}x} &\sim 1 + k\sqrt{\frac{\pi}{2}} \cos\left(\delta
\phi+\frac{\pi}{4}\right) \nonumber \\
&-\frac{\sqrt{\pi}}{4}\, k^2\left[1+(2-\sqrt{2})\cos\left(2 \delta\phi
+\frac{\pi}{4}\right)\right]\,
\end{align}
for $x\gg1$. We note that the leading-order-in-$k$ correction to the asymptotic
gradient in the phase equation and the leading-order correction to $b$ in the
frequency equation can be made to vanish for $\delta\phi=\pi/4$. However, even
with this choice, the next-to-leading-order correction does not vanish, and so
although the effect of the resonance can be suppressed for certain
values of $\delta\phi$, it cannot be eliminated.

Let us now relax assumption (c); that is, let us include higher $\ell$ modes in
the resonant differential equations. Consider first an evolution equation
of the form
\be
\frac{{\rm d}y}{{\rm d}x} = 1 + k \cos(\ell \,xy)\,,
\ee
where $n$ is some integer. By writing $Y=\sqrt{\ell}y$, $X=\sqrt{\ell} x$, we can
rewrite this equation as
\be
\frac{{\rm d}Y}{{\rm d}X} = 1 + k \cos(XY),
\ee
reducing it to the same form that we considered before. We note in
particular that since the solution to this equation behaves as $Y \sim (1+c)
X$ for $x \gg 1$, this implies that $y \sim (1+c) x$ and hence the value of $c$ is unchanged as a function of $k$. The value of $b$ would be modified, however.

In the phase case, if we modify the equation to
\be
\frac{{\rm d}^2\phi}{{\rm d}x^2} = 1 + k \cos(\ell \, \phi)\,
\ee
with $n$ again an integer, the substitution $\Phi=\ell\phi$, $X=\sqrt{\ell} x$ gives
\be
\frac{{\rm d}^2\Phi}{{\rm d}X^2}= 1 + k \cos(\Phi).
\ee
Using the solution to this equation that we found earlier, we find that $\Phi
\approx X^2/2 + \alpha(k) X + \cdots$ and we deduce that
\be
\phi \approx \frac{1}{2} x^2 + \frac{\alpha(k)}{\sqrt{\ell}} X + \cdots\,.
\ee
Clearly then, higher-$\ell$ modes in the sum of \eqref{probdef} are suppressed
by a factor of $1/\sqrt{\ell}$. 

Let us now consider the case where we have more than one term on the righthand side of the
evolution equation. In the frequency resonance case, we would have an equation like
\be
\frac{{\rm d}y}{{\rm d}x}=1+k_1\cos(xy)+k_\ell\cos(\ell\,xy).
\ee
To solve this equation, we can proceed as before, making an ansatz of the form
\begin{align}
y_1 &\equiv (1+c)x + b + \frac{a^{(1)}_1}{x} \sin[(1+c)x^2 + bx ] \nonumber \\
&+ \frac{a^{(\ell)}_1}{x} \sin[\ell ((1+c)x^2 + bx)].
\end{align}
Inserting this into the differential equation, we find the same source terms
involving $\sin^2(\cdot)$ that we found earlier, one from each mode. We also
find various cross source terms, but these do not contribute to the
zero-frequency part of the solution because they are products of oscillatory
functions with unequal frequencies. We deduce then that the solution for $c$ at
leading order is the sum of the solutions treating each of the modes
individually. At the next order, cross terms will come in that may be important,
but these will be subdominant and further exploration of these is beyond the
scope of this paper.

In the case of the phase equation we would have
\be
\frac{{\rm d}^2\phi}{{\rm d}x^2}=1+k_1\cos(\phi)+k_\ell\cos(\ell\phi).
\ee
The linear-in-$k$ term in the solution of this equation can readily be seen to
be the sum of the linear-in-$k$ solutions to the equation with only one of the
cosine terms on the righthand side. Cross terms again come in at higher order, but
by the preceding argument the size of the corrections from the $k_\ell$ term are
suppressed relative to the dominant mode by factors of $1/\sqrt{\ell}$. 

\section{Conclusions}
\label{conclusions}

We have studied the behavior of the solution to two differential
equations that describe the gravitational-wave phase and frequency evolution
during an EMRI that experiences a resonant transition. We have found two general
differential equations that might describe this behavior: one based on the assumption that the coefficients of a Fourier expansion of the self-force in frequency are continuous at a resonance, and the other based on the assumption that it is the Fourier phase coefficients that are continuous at resonance. We have solved both differential equations at late times using asymptotic methods. 

Depending on the strength of the resonance (controlled by the parameter $k$),
both resonant equations lead to solutions with qualitative different behavior:
Weak resonances ($k<1$) lead to linear temporal growth of the frequency and
quadratic growth of the orbital phase; strong resonances ($k>1$) lead to linear
temporal decay of the frequency function, leading to a constant-phase offset at
late times. 

Even though the differential equations for phase and frequency resonances might
both describe the evolution of an EMRI through a resonance, we have found that the
evolution depends on which differential equation one assumes. That is,
at late times and in the weak resonance case, the evolution of the frequency in the phase and frequency resonant cases possess different asymptotic slopes. The difference in slope depends on the value of $k$,
with the frequency evolution acquiring a $k$-dependent memory in the frequency resonant case that is absent for phase resonances. Further work is required to explore which of the two equations is in fact most applicable to the
EMRI resonance problem.

We also studied the transition between weak and strong resonances. We found that
at the transition point $k=1$, there is a square-root branch cut in the solution to the
frequency resonance equation. Close to this point, we proved that frequency solutions
bunch up into families that decay as $1/x$ exponentially fast. In fact, there is
an alternation between stable (bunching up of solutions) and unstable behavior,
depending on the branch of solutions considered. 

Future work should concentrate on exploring which of these equations is applicable to EMRI evolutions in practice and what the implications are for the construction of waveform template models of EMRI signals. The existence of a memory effect in the frequency resonances is particularly interesting and would have a profound impact on our ability to detect EMRI signals.
An approximate post-Newtonian prescription for the self-force on resonance has been suggested~\cite{Flanagan:2010cd} and would provide a suitable framework in which to explore these questions further. Whichever of the two equations applies to the problem that motivated this work, the results described in this paper provide important insights into the behaviour of the solutions to these differential equations and predictions for the change in the frequency and phase of the evolution as the orbit passes through a resonance. These results will be invaluable for constructing approximate models to describe the evolution of EMRI orbits.

\acknowledgments

CMB is supported by grants from the Leverhulme Foundation and the
U.S.~Department of Energy. JG's work is supported by the Royal Society. NY
acknowledges support from NASA grant NNX11AI49G, under sub-award 00001944 and
NASA through the Einstein Postdoctoral Fellowship Award Number PF0-110080 issued
by the Chandra X-ray Observatory Center, which is operated by the Smithsonian
Astrophysical Observatory for and on behalf of NASA under contract NAS8-03060.
JG thanks the MIT Kavli Institute for Astrophysics and NY thanks the Yukawa
Institute for Theoretical Physics for their hospitality while this paper was
being finished. We also thank Scott Hughes for useful discussions.


%

\end{document}